\documentstyle[preprint,aps]{revtex}
\begin{document}
\draft
\title {CVM analysis of crossover in the semi--infinite Ising model}
\author {Alessandro Pelizzola}
\address {Dipartimento di Fisica and Unit\'a INFM,
Politecnico di Torino, I-10129 Torino, Italy}
\maketitle
\begin{abstract}
The crossover behavior of the
semi--infinite three dimensional Ising model
is investigated by means of Pad\'e approximant analysis of
cluster variation method results.
We give estimates for ordinary critical as well as for multicritical
exponents, which are in very good agreement with extensive Monte Carlo
simulations.
\end{abstract}

\vfill

\def\npar{\par \vskip -.3 truecm \noindent}


\bigskip

\noindent \underline{Address:} \npar
Alessandro Pelizzola \npar
Dipartimento di Fisica \npar
Politecnico di Torino \npar
Corso Duca degli Abruzzi 24 \npar
I-10129 Torino, ITALY \npar
Tel. +39-11-5647318 \npar
Fax. +39-11-5647399 \npar
E-mail: pescarmona@polito.it \par

\vfill\eject

\def\js{J_{\rm s}}

The present paper reports on a preliminary
investigation of the multicritical behavior of the
semi--infinite three dimensional Ising model by
means of a recently proposed technique
\cite{cvmpade} which combines the off--critical
great accuracy of the cluster variation method (CVM)
\cite{cvm1,cvm2,cvm3} with the analysis tools for critical
behavior based on the well--known Pad\'e
approximants \cite{guttmann}.

The semi--infinite Ising model (see Ref.\ \cite{binder} for a review)
is described by the (reduced) hamiltonian
\begin{equation}
\beta H = - \js \sum_{\langle i j \rangle \subset
\partial \Lambda} s_i s_j - J \sum_{\langle k l
\rangle \not \subset \partial \Lambda} s_k s_l,
\end{equation}
where $s_i = \pm 1$ is an Ising spin at the
site $i$ of the lattice $\Lambda$, $\partial
\Lambda$ is the free surface of the lattice,
$\js, J > 0$ are the surface and bulk nearest
neighbor interaction strengths, respectively and $\beta = 1/k_B T$, with
$k_B$ Boltzmann's constant and $T$ absolute temperature.
In the following, we will take $\Lambda$ to be a
simple cubic lattice and $\partial \Lambda$ a
(100) free surface.

The phase diagram of this model
is well--known. For $\Delta \equiv \js/J - 1 <
\Delta_c$ we have the so--called {\it ordinary}
transition at the bulk critical temperature,
with simultaneous (but with different critical
exponents) disordering of bulk and surface.
For $\Delta > \Delta_c$ we have first the {\it
extraordinary} transition at the bulk critical
temperature, at which only the bulk disorders,
and then the {\it surface} transition, at a
higher temperature, at which the surface
disorders. For $\Delta = \Delta_c$ we have a
multicritical point at the bulk critical
temperature, at which the {\it special} transition
takes place. The {\it surface} transition is in
the two dimensional Ising universality class
and, in the vicinity of the special point, its
critical temperature has the power--law behavior
\begin{equation}
T_{\rm cs} - T_{\rm cb} \sim (\Delta - \Delta_c)^\phi,
\label{tcs}
\end{equation}
where $\phi$ is the crossover exponent associated
to the multicritical point.

Our purpose is to show that reliable estimates
for the location of the multicritical point and
the crossover exponent, as well as other critical
exponents, can be obtained by means of a new
technique \cite{cvmpade} which relies on a
Pad\'e approximant analysis of CVM results.
The basic idea is to find a region, far enough
from the critical point, where the quantity of
interest can be obtained with a given accuracy
(say, $10^{-5}$) from the CVM, and then to construct
$[L,M]$ Pad\'e approximants for functions like
that in Eq.\ \ref{beta1star} on a suitably chosen
set of $L + M + 1$ points in the region of accuracy
(see Ref.\ \cite{cvmpade} for more details).

In what follows, we shall use the $4 \times N$
approximation which we have introduced in Ref.\
\cite{mycvm}. Here the semi--infinite lattice is
approximated by a film of $N$ layers, with the
topmost layer representing the free surface,
and the bottom layer constrained to the bulk, which
in turn is studied in the cube approximation. In
this system the maximal clusters for the CVM are
chosen as those clusters with $4N$ sites, formed by
a column of $N-1$ elementary cubes. The
approximation is formally similar to the square
(Kramers--Wannier) approximation for the square
lattice and in fact reduces to it for vanishing
bulk interaction.

By means of the $4 \times N$ CVM approximation with
$N = 4$ (but we had to use also $N = 5$ for $\Delta
< 0$ and low $J$), we determined the behavior of
the surface layer magnetization $m_1$ as a function
of the temperature for $J > J_{\rm min} = 0.30$ and
several values of $\Delta$, in the range $-0.50 \le
\Delta \le 0.60$. In this region the maximum error
on $m_1$ is estimated \cite{cvmpade,mycvm} to be
about $10^{-5}$ or less.

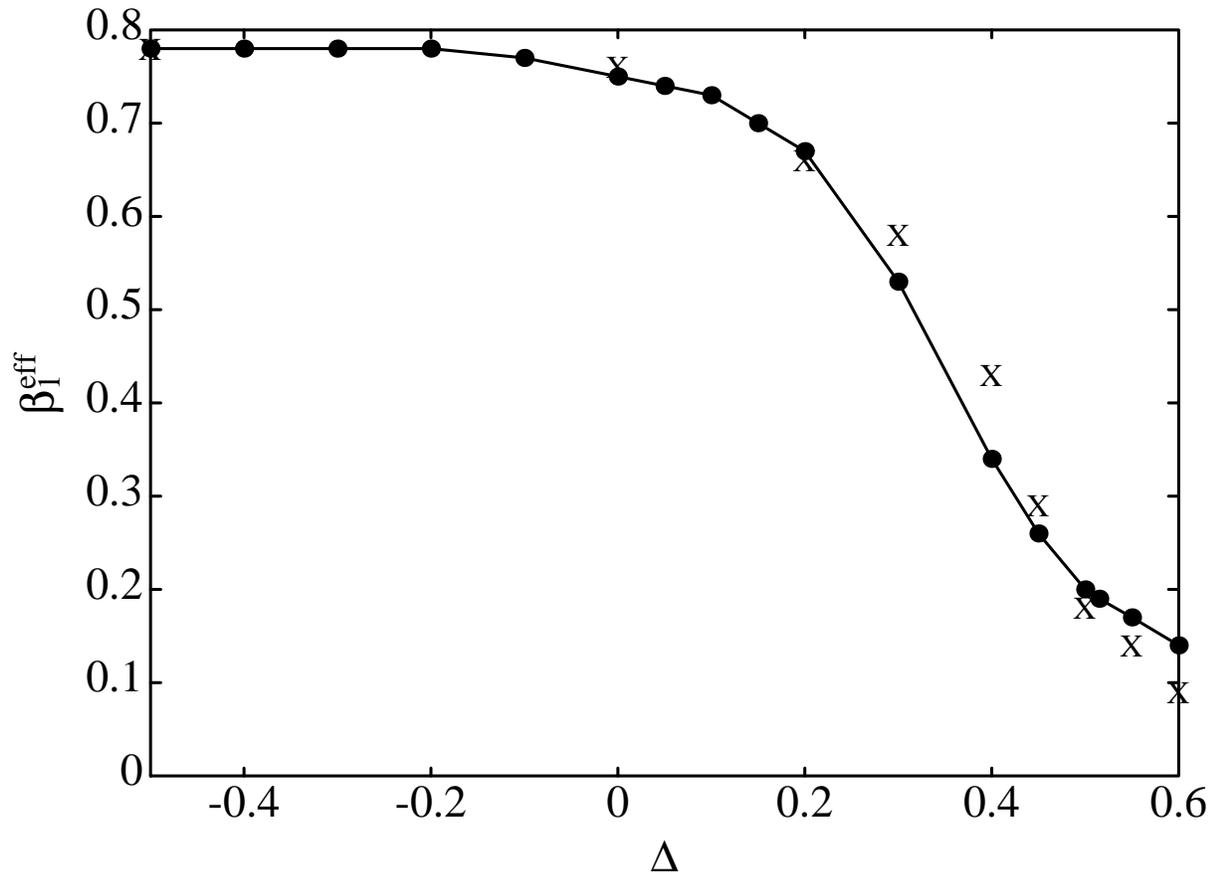
\begin{figure}
\caption{Effective exponent for the surface layer magnetization. Present
work ($\bullet$, solid line is a guide to the eye)
and Monte Carlo ($\times$, from Ref.\ [9]).}
\label{beta1eff}
\end{figure}

We then determined, following the procedure proposed
in Ref.\ \cite{cvmpade}, Pad\'e approximants for the
function
\begin{equation}
\beta_1^*(z,\Delta) = (z_{\rm c} - z)\frac{d}{dz} \ln m_1(z,\Delta),
\label{beta1star}
\end{equation}
where $z = e^{-J}$ and $z_{\rm c} \simeq 0.801$
\cite{cvmpade,mcising}. We obtained very stable Pad\'e
tables for $\beta_1^{\rm eff}(\Delta) =
\beta_1^*(z_{\rm c},\Delta)$, and our estimates are
plotted in Fig.\ \ref{beta1eff}, which shows very
good agreement with available Monte Carlo data
\cite{binlan}. As already mentioned in Ref.\
\cite{cvmpade} the exponent for the ordinary transition
is estimated as $\beta_1 \simeq 0.78$,
again in very good agreement with Monte Carlo
estimates \cite{binlan,rdww}.

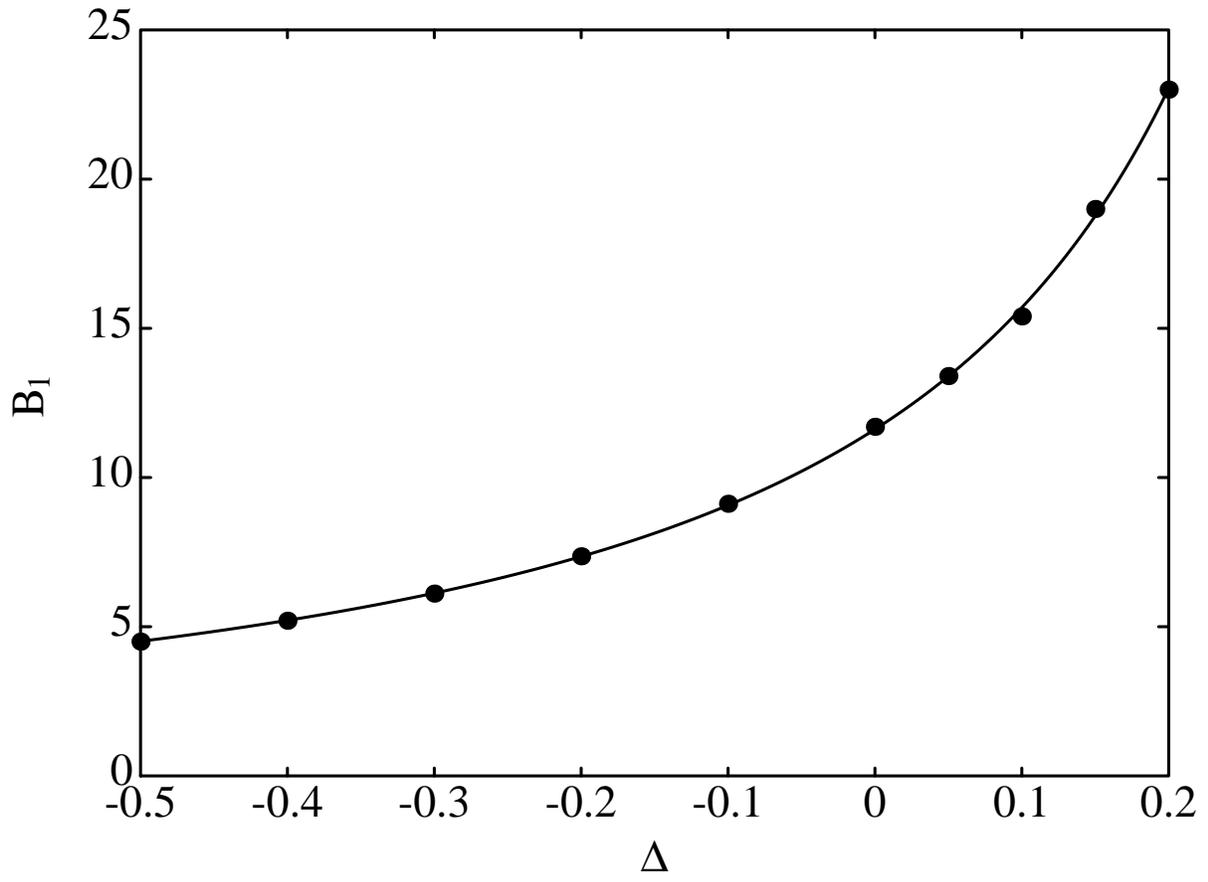
\begin{figure}
\caption{Critical amplitude of the surface layer magnetization. The solid
line is our fit.}
\label{b1fit}
\end{figure}

For $\Delta > 0$, the effective exponent is
appreciably different from its limiting value
$\beta_1$, indicating that the crossover
region is quite large. This allows us to study
the multicritical behavior of our model by
analyzing the behavior of the critical
amplitude $B_1(\Delta)$ defined by
\begin{equation}
m_1(z,\Delta) \sim B_1(\Delta) (z_{\rm c} - z)^
{\beta_1}, \qquad z \to z_{\rm c}^{-}.
\end{equation}
In fact, since near the multicritical point one has
\begin{equation}
m_1(z,\Delta) \simeq (z_{\rm c} - z)^
{\beta_1^{\rm m}} \tilde m_1(x), \qquad x = \frac
{\Delta_{\rm c} - \Delta}{(z_{\rm c} - z)^\phi},
\end{equation}
the critical amplitude $B_1(\Delta)$ must diverge as
$(\Delta_{\rm c} - \Delta)^{-(\beta_1 -
\beta_1^{\rm m})/\phi}$. By fitting this
behavior we found
$\Delta_c \simeq 0.515$  and $(\beta_1 -
\beta_1^{\rm m})/\phi \simeq 1.39$, which, using
$\beta_1^{\rm m} = \beta_1^{\rm eff}(\Delta_{\rm c})
\simeq 0.19$, yields $\phi \simeq 0.42$. The calculated
amplitudes, together with the theoretical curve, are shown
in Fig.\ \ref{b1fit}, where it can be seen that the fit
is remarkably accurate.

By comparing these results with those from extensive
Monte Carlo simulations in Tab.\ 1
we see that our values for $\Delta_{\rm c}$ and
$\beta_1^{\rm m}$ are again in very good agreement
with previous results, while for the crossover
exponent $\phi$, where the situation is less clear,
we obtained a value which turns out to be closer to
Ruge et al.\ result than to Binder and Landau one.
As a general remark, our technique seems to be
fairly more accurate than the second order
$\epsilon$--expansion. It is also quite simple and does not require very
large CPU times ($\sim 25$ hours on a DEC Alpha for this work).

\begin{table}
\caption{ Comparison with Monte Carlo simulations and $\epsilon$--expansion.
\label{table1}}
\begin{tabular}{lccc}
& $\Delta_{\rm c}$ & $\beta_1^{\rm m}$ & $\phi$ \\
\tableline
Present work               & 0.515  & 0.19  & 0.42  \\
Landau and Binder [9]      & 0.52   & 0.18  & 0.461 \\
Ruge et al. [10]           & 0.5004 & 0.237 & 0.59  \\
$\epsilon$--expansion [11] &   --   & 0.245 & 0.68  \\
\end{tabular}
\end{table}

\begin{references}
\bibitem{cvmpade} A. Pelizzola, Phys. Rev. E 49 (1994) R2503.
\bibitem{cvm1} R. Kikuchi, Phys. Rev. 81 (1951) 988.
\bibitem{cvm2} G. An, J. Stat. Phys. 52 (1988) 727.
\bibitem{cvm3} T. Morita, J. Stat. Phys. 59 (1990) 819.
\bibitem{guttmann} A.J. Guttmann, in {\it Phase transition and
Critical Phenomena}, vol. 13, edited
by C. Domb and J.L. Lebowitz (Academic,
London, 1989) and refs. therein.
\bibitem{binder} K. Binder, in {\it Phase transition and
Critical Phenomena}, vol. 8, edited
by C. Domb and J.L. Lebowitz (Academic,
London, 1983).
\bibitem{mycvm} A. Pelizzola, submitted to Physica A.
\bibitem{mcising} A.M. Ferrenberg and
D.P. Landau, Phys. Rev. B 44 (1991) 5081.
\bibitem{binlan} D.P. Landau and
K. Binder, Phys. Rev. B 41 (1990) 4633.
\bibitem{rdww} C. Ruge et al., J. Stat. Phys. 73 (1993) 293.
\bibitem{epsilon} H.W. Diehl and
S. Dietrich, Z. Phys. B 42 (1981) 65.
\end{references}
\end{document}